\documentclass[a4paper,11pt]{article}
\usepackage{pos}
\usepackage{subcaption}

\usepackage{lineno}

\let\OLDthebibliography\thebibliography
\renewcommand\thebibliography[1]{
  \OLDthebibliography{#1}
  \setlength{\parskip}{0pt}
  \setlength{\itemsep}{0pt plus 0.01ex}
}

\title{Physics program and performance of the ALICE Forward Calorimeter upgrade (FoCal)}

\author*[a]{Laura Huhta}
\onbehalf{on behalf of the ALICE collaboration}

\affiliation[a]{Department of Physics, University of Jyväskylä,\\
 P.O. Box 35, FI-40014, Jyväskylä, Finland}

\emailAdd{laura.maria.huhta@cern.ch}

\abstract{The FoCal is a high-granularity forward calorimeter to be installed as an ALICE upgrade subsystem during the LHC Long Shutdown 3 and take data during the LHC Run 4. 
The FoCal detector, covering a pseudorapidity interval of $3.2 < \eta < 5.8$, extends the ALICE physics program with the capability to investigate gluon Parton Distribution Functions (PDFs) in the low-$x$ regime, down to $x \approx 10^{-6}$. 
FoCal measurements will provide experimental constraints for PDFs in a region of phasespace where experimental data is scarce, as well as enable the study of non-linear QCD effects like gluon saturation.

The FoCal detector consists of two components. The highly-granular Si+W electromagnetic calorimeter (FoCal-E) with pad and pixel longitudinal and transverse segmented readout layers provides high spatial resolution for discriminating between isolated photons and decay photon pairs.
The hadronic calorimeter (FoCal-H) is constructed from copper capillary tubes filled with scintillator fibers and is used for isolation energy measurement and jets.
With this detector design, FoCal is capable of measuring direct photons, jets, and the photo-production of vector mesons such as the \jpsi \,in $\mbox{p--Pb}$ and $\mbox{Pb--Pb}$ ultra-peripheral collisions. In addition, correlations of different probes can be studied, including $\gamma$--jet, jet--jet and \piz -- \piz correlations. These measurements will place stringent constraints to various theoretical models incorporating non-linear QCD effects. 

This contribution gives an overview of the FoCal physics program. Also, recent experimental results of ever-improving prototypes of the detector, which were tested at the Test Beam facilities of CERN in the years 2021--2023, as well as simulation studies showcasing the robustness of the detector design and its physics potential are presented.}

\FullConference{31st International Workshop on Deep Inelastic Scattering (DIS2024)\\
 8–12 April 2024\\
Grenoble, France\\}

\begin{document}
\maketitle

\section{Introduction}

Due to the non-Abelian nature of Quantum Chromodynamics (QCD), it is expected that, at low Bjorken-$x$, QCD evolution is non-linear and the gluon density of matter saturates. To explore these phenomena, the Forward Calorimeter (FoCal) upgrade is to be installed in the ALICE experiment at the Large Hadron Collider (LHC) at CERN for data-taking in Run 4 (2029-2032) \cite{LOI}. FoCal will cover a pseudorapidity interval of $3.2 < \eta < 5.8$, extending the capability of ALICE to explore QCD at unprecedentedly low values of $x\approx 10^{-6}$. 

In order to probe the low-$x$ nature of QCD evolution, FoCal will provide forward measurements of direct photons, neutral mesons, vector mesons, jets, Z-bosons, and their correlations in hadronic and ultra-peripheral $\pPb$ and $\PbPb$ collisions \cite{Phys}. The FoCal design is optimized for this physics program. The detector consists of a Si+W electromagnetic calorimeter, composed of two high-granularity pixel layers of cell size $\approx 30 \times 30\ \mathrm{\mu m}^2$ and 18 layers of silicon pad sensors, with transverse cell size of $1\ \mathrm{cm}^2$, and a spaghetti-type sampling hadronic calorimeter structured of scintillation fibers inside copper tubes. 

The forward measurement of direct photons at low momentum $p_{\rm T}$ imposes a strict constraint on detector performance, as it requires excellent discrimination of single-photon showers from merged electromagnetic showers caused by decaying neutral mesons. This is achieved with the excellent spatial resolution provided by the FoCal-E. Additional discrimination is provided by good FoCal-H energy resolution and linearity. These requirements also result in good performance for jets, neutral mesons, quarkonia, \zboson\ bosons, and other observables \cite{perf}.

\section{Physics performance}

The physics performance of FoCal is assessed using $\pp$ and $\pPb$ collision events at $\sqrtS = 14 $ TeV and $\sqrtSnn = 8.8$ TeV, respectively, simulated using PYTHIA8 \cite{pythia}, HIJING \cite{hijing}, and STARlight \cite{starlight} in the case of ultra-peripheral collisions (UPCs), propagated through the idealized FoCal detector using GEANT3 \cite{geant}. To save computational cost, we consider a sandwich-type FoCal-H geometry \cite{LOI} which does not have a large effect on the performance results \cite{perf}. At the moment no electronics response effects or triggers were implemented in the simulations. Implementation of these effects is currently
in progress, but is not expected to affect the performance assessment significantly. Studies have been performed \cite{perf} on a wide set of experimental observables, of which some results will be presented in this section.
 
\subsection{Prompt photon and neutral meson measurements}

\begin{figure}[]
    \begin{subfigure}{.4\textwidth}
        \centering
        \includegraphics[width=0.79\textwidth]{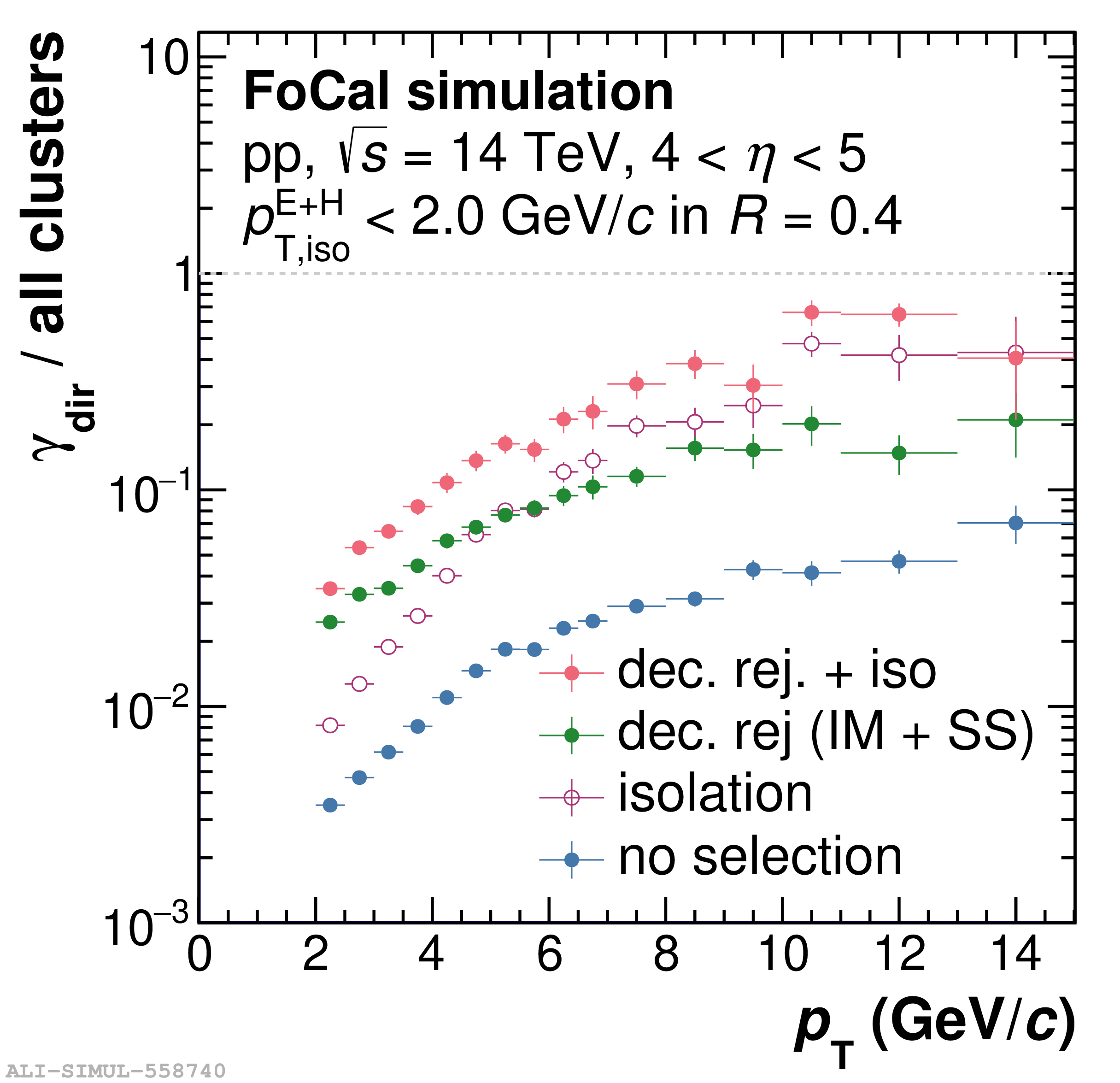}
    \end{subfigure}%
    \begin{subfigure}{.6\textwidth}
        \centering
        \includegraphics[width=0.79\textwidth]{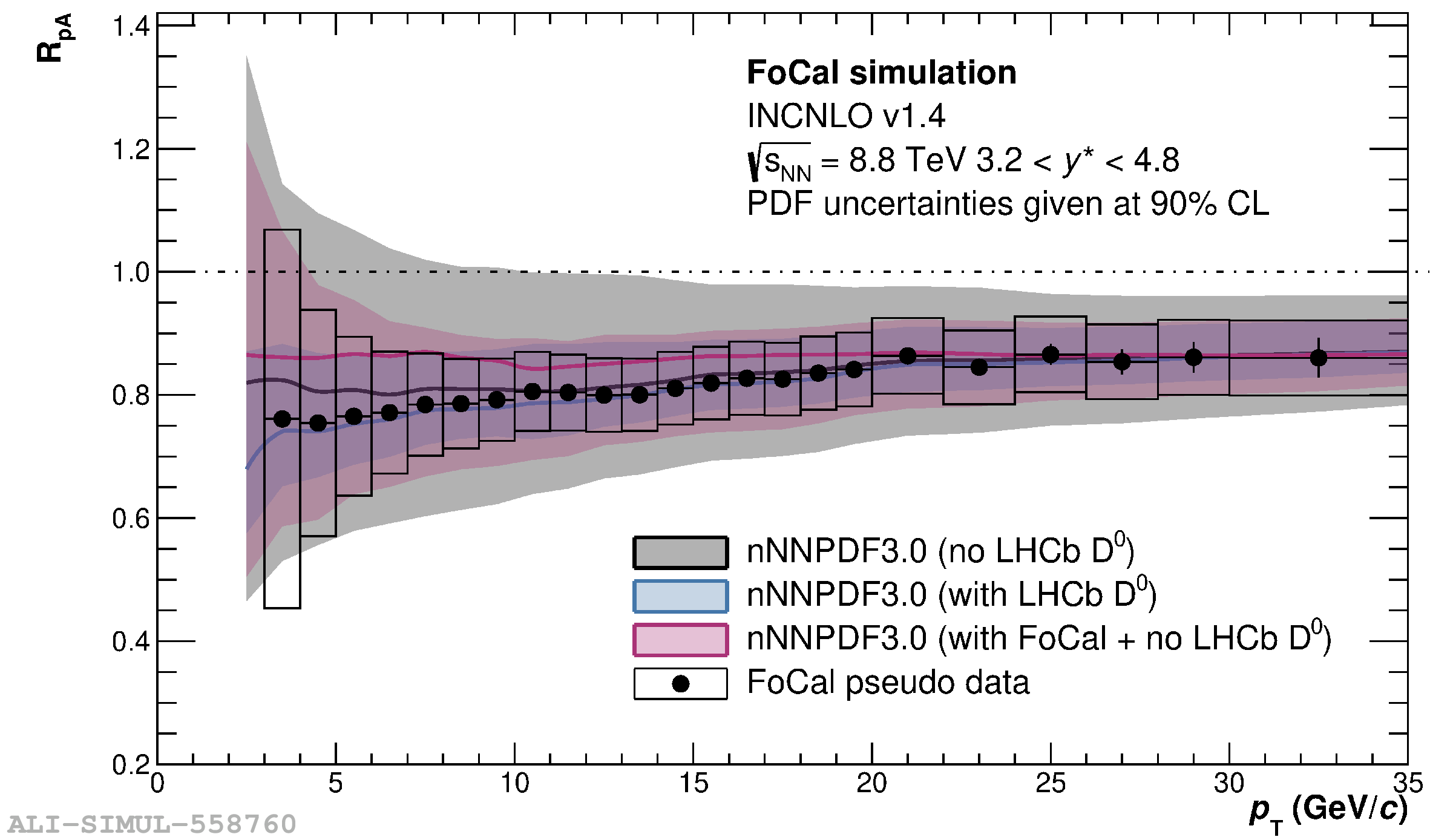}
    \end{subfigure}
    \caption{\textbf{Left:} Prompt photon signal fraction as a function of cluster momentum $p_{\rm T}$. \textbf{Right:} Distribution of the nuclear modification factor $R_\mathrm{pA}$ for $\pPb$ collisions, for prompt photons obtained from the FoCal pseudodata and from QCD calculations using nNNPDF3.0 with and without LHCb D-meson measurements and FoCal pseudodata \cite{perf}.}
    \label{fig:photons}
\end{figure}

Direct photons consist of prompt and fragmentation photons. 
At forward rapidities, experimentally measured isolated photons are dominated by the QCD Compton process, $\mathrm{qg}\rightarrow \photon \mathrm{q}$, providing access to gluon distributions in the lowest order of perturbation theory.
Prompt photon contributions can be enriched using various kinematic and selection cuts. 
First, a selection on isolation energy is performed. Then, decay photon rejection is performed in two steps, first based on the shower shape, then based on invariant mass tagging of photon pairs originating from neutral pions. As shown in Figure \ref{fig:photons} (left), these steps improve the signal fraction by about a factor of ten.

The inclusion of FoCal prompt photon data in global PDF fits will provide additional insight into factorization and universality in nuclear environments. In Figure \ref{fig:photons} (right), FoCal pseudodata of the nuclear modification factor $\RpPb$ for prompt photons is shown. The central values of the FoCal pseudo-data are based on NLO calculations using the INCNLO program. Also shown are the $\RpPb$ distributions obtained using nNNPDF3.0 \cite{nnnn} reweighted with the FoCal pseudodata or LHCb measurements of D-meson production \cite{lhcb}. The two separate probes differ in their sensitivity to final state effects, therefore, the inclusion of both in PDF fits offers a multimessenger approach to explore the low-$x$ region.

Neutral mesons decaying fully into photons or electrons can be reconstructed using electromagnetic showers in the FoCal-E. The most abundant of these are $\pi^0 $, $\eta$ and $\omega$. Vector mesons decaying via di-electrons can also be reconstructed. FoCal simulation studies have shown good single pion reconstruction efficiency, as well as the capability to measure both $\eta$ and $\omega$ mesons \cite{perf}. Clusterization parameters can be tuned in order to obtain a better performance in certain kinematical regions, such as high $\pi^0$ energy.

Photon and hadron-triggered correlations in the forward region of $\pPb$ collisions also probe small-$x$ gluon dynamics, but with a different sensitivity than inclusive yields. Correlated yield suppression probes the gluon density, similarly to measurements of inclusive production, while the measurement of angular decorrelation is in addition sensitive to the coherence of the gluonic wavefunction. 
We have developed a sideband subtraction method to measure forward \piz--\piz correlations and made first feasibility studies on $\gamma$--\piz correlations \cite{perf}. Correlation studies can be significantly extended both to jets and correlations over a large rapidity range.

\subsection{Jet measurements}

Forward inclusive jets, photon-jet correlations and dijets are sensitive observables to gluon saturation. Especially dijets can be studied via their momentum imbalance $k_\mathrm{T}$, which probes the saturation scale $Q_\mathrm{sat}$. 
 Jet reconstruction is performed at both the particle and detector level utilizing the anti-$k_\mathrm{T}$ clustering algorithm with E-scheme recombination. For jets of radius $R = 0.6$, the pseudorapidity acceptance is $4.0 < \eta_\mathrm{jet} < 4.9$. No cluster or cell energy threshold cut is applied. 
 
\begin{figure}[]
    \centering
    \begin{subfigure}{.5\textwidth}
      \centering
      \includegraphics[width=.89\textwidth]{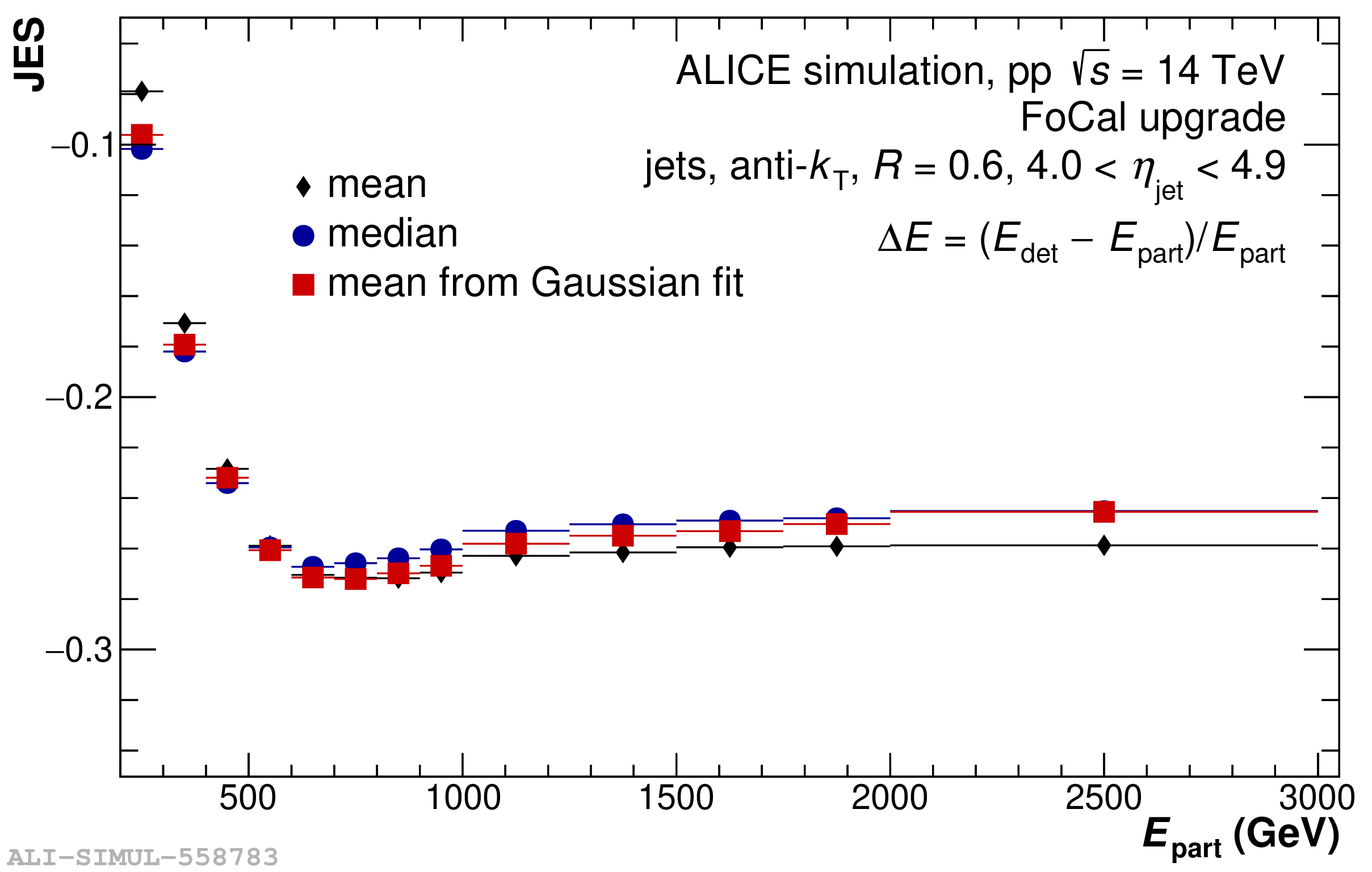}
      \label{fig:jes}
    \end{subfigure}%
    \begin{subfigure}{.5\textwidth}
      \centering
      \includegraphics[width=.89\textwidth]{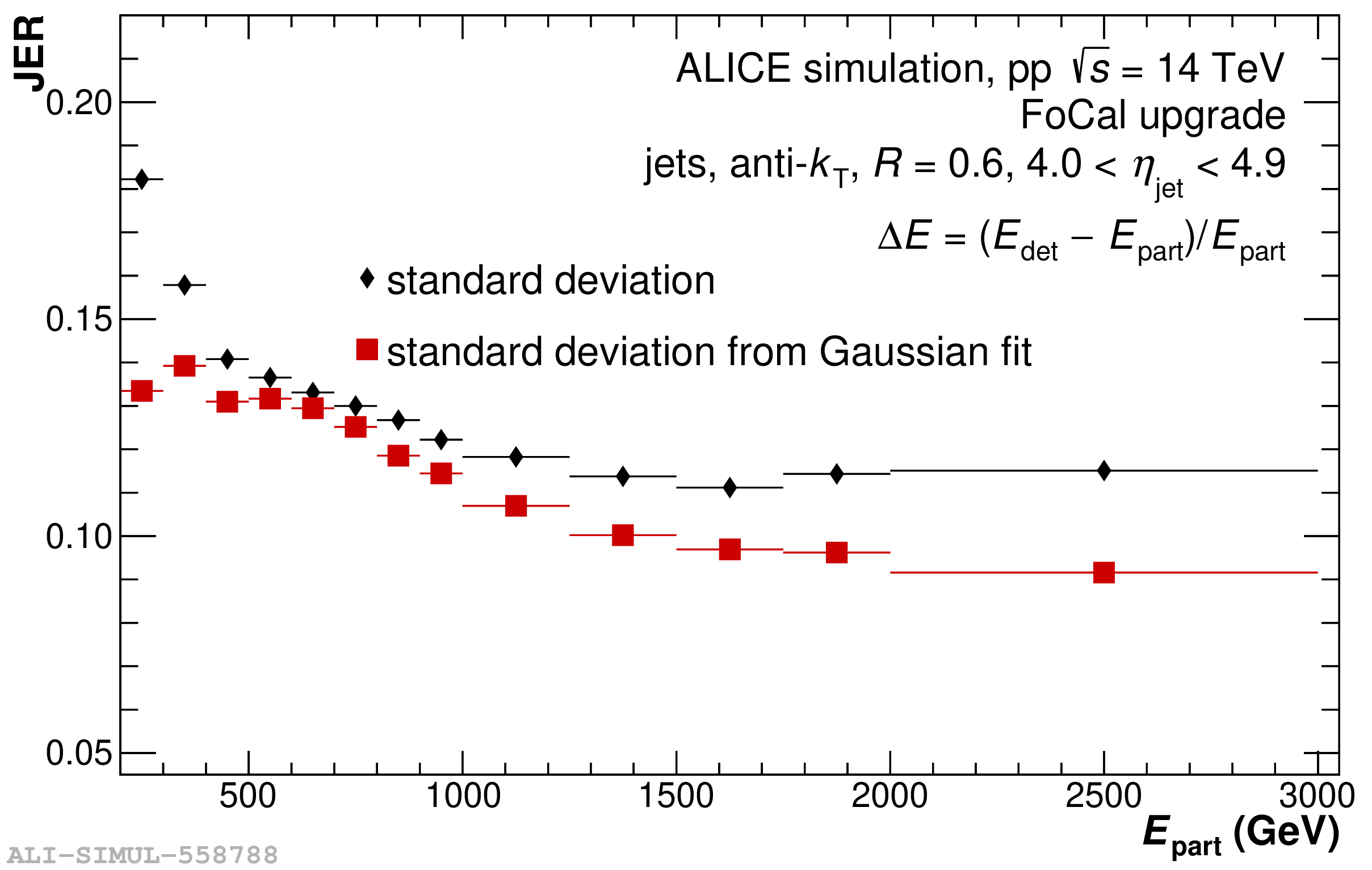}
      \label{fig:jer}
    \end{subfigure}
    \caption{The Jet Energy Scale (left) and Jet Energy Resolution (right) for anti-$k_\mathrm{T}$ jets with jet resolution parameter $R = 0.6$, calculated from the $\Delta E$ distribution.}
    \label{fig:jets}
\end{figure}

 Jet reconstruction performance is quantified by the relative energy difference $\Delta E$ of matched jets at the detector and particle levels. The Jet Energy Scale (JES) is characterized by the mean of the $\Delta E$ distributions, while the Jet
Energy Resolution (JER) is characterized by its RMS. 
These distributions are shown in Figure \ref{fig:jets} for varying particle level jet energy. The JER is better than 15 \%, while the JES shows energy loss up to 28 \%. This can be attributed to the geometric considerations of jets at forward rapidities, where the transverse size of the hadronic shower in FoCal-H is larger than the jet radius $R = 0.6$. Additional studies also suggest that the JES can be improved by biasing on the neutral energy fraction of the jets \cite{perf}.

\subsection{Vector meson photoproduction in ultra-peripheral collisions}

At leading order in pQCD, the photoproduction cross sections of heavy vector mesons are proportional to the gluon density in the target nucleon/nucleus squared. Thus, UPCs can be useful in constraining PDFs and studying the non-linearity of gluon density at high center-of-mass energies, $W_{\photon p}$, of the emitted photon and the proton projectile.

\begin{figure}[]
    \centering
    \begin{subfigure}{.55\textwidth}
      \centering
      \includegraphics[width=.79\textwidth]{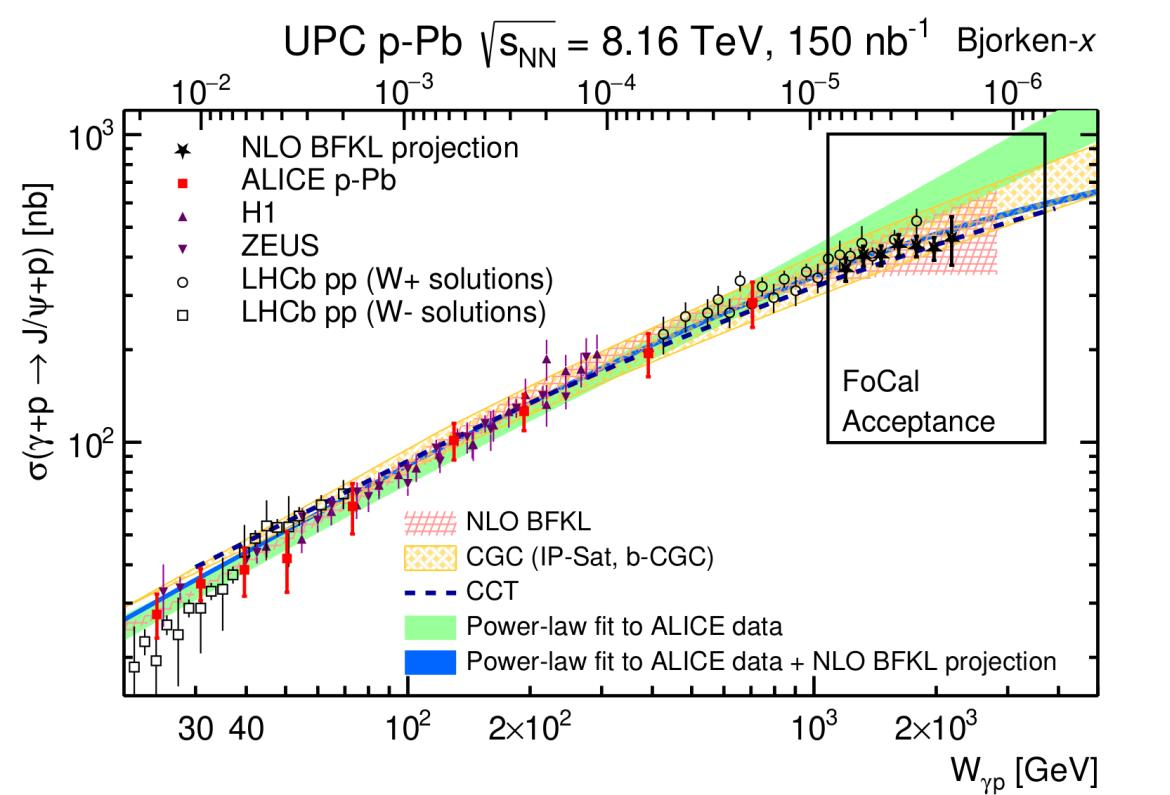}
    \end{subfigure}%
    \begin{subfigure}{.45\textwidth}
      \centering
      \includegraphics[width=.79\textwidth]{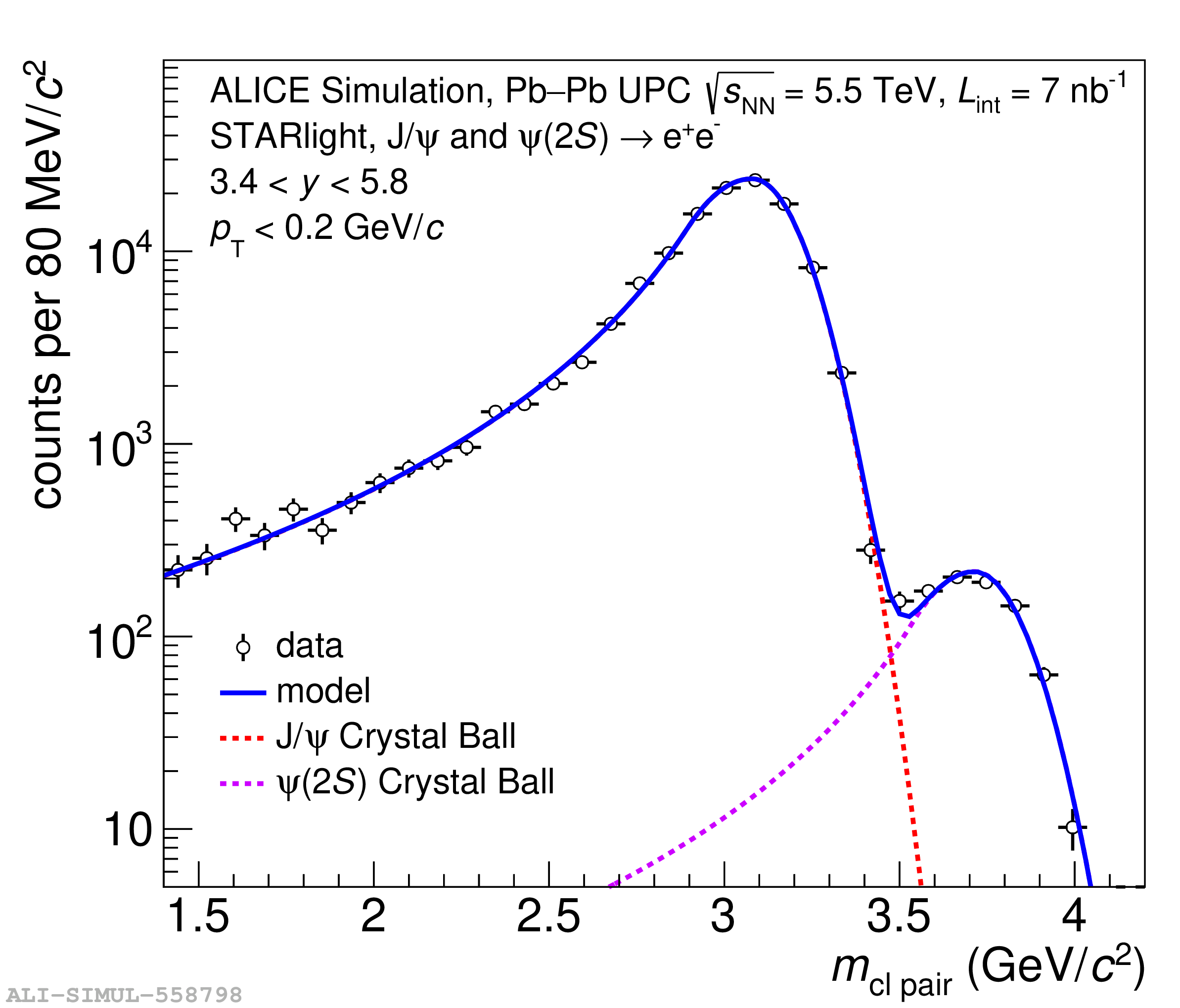}
    \end{subfigure}
    \caption{ \textbf{Left:}  $\jpsi$ photoproduction cross section in $\pPb$ UPCs for data, theoretical predictions and expected FoCal coverage \cite{BNT}. \textbf{Right:} Invariant mass distribution of  $\jpsi$ and $\psi (2S)$ candidates based on STARLight + GEANT simulations \cite{perf}.}
    \label{fig:UPC1}
\end{figure}

FoCal provides a unique kinematic coverage, extending significantly the kinematic reach of current ALICE measurements and complementing the program foreseen at the Electron-Ion Collider (EIC). FoCal provides access to unprecedentedly low values of $x$, extending existing measurements up to $W_{\photon p} \approx  2\, \mathrm{TeV}$ (and down to 10 GeV) in $\pPb$ ($\Pbp$) and $\PbPb$ collisions, which is shown in Figure \ref{fig:UPC1} (left). In Figure \ref{fig:UPC1} (right), the successful reconstruction of $\jpsi$ and $\psi (2S)$ from signals simulated using STARLight and GEANT is shown. 
 
\section{Test beam results}

\begin{figure}
    \centering
    \begin{subfigure}{.42\textwidth}
      \centering
      \includegraphics[width=.84\textwidth]{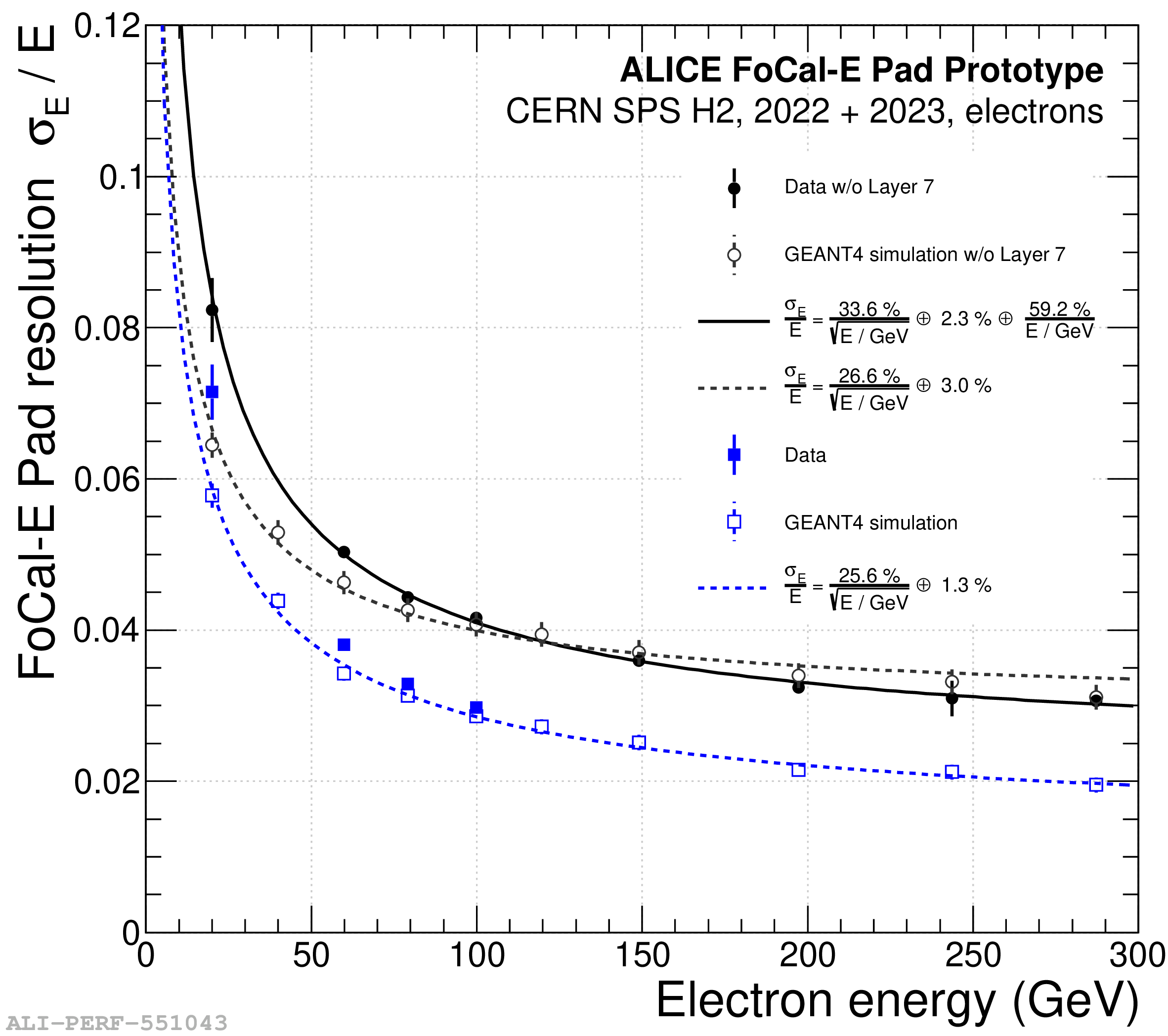}
    \end{subfigure}%
    \begin{subfigure}{.58\textwidth}
        \centering
        \includegraphics[width=.89\textwidth]{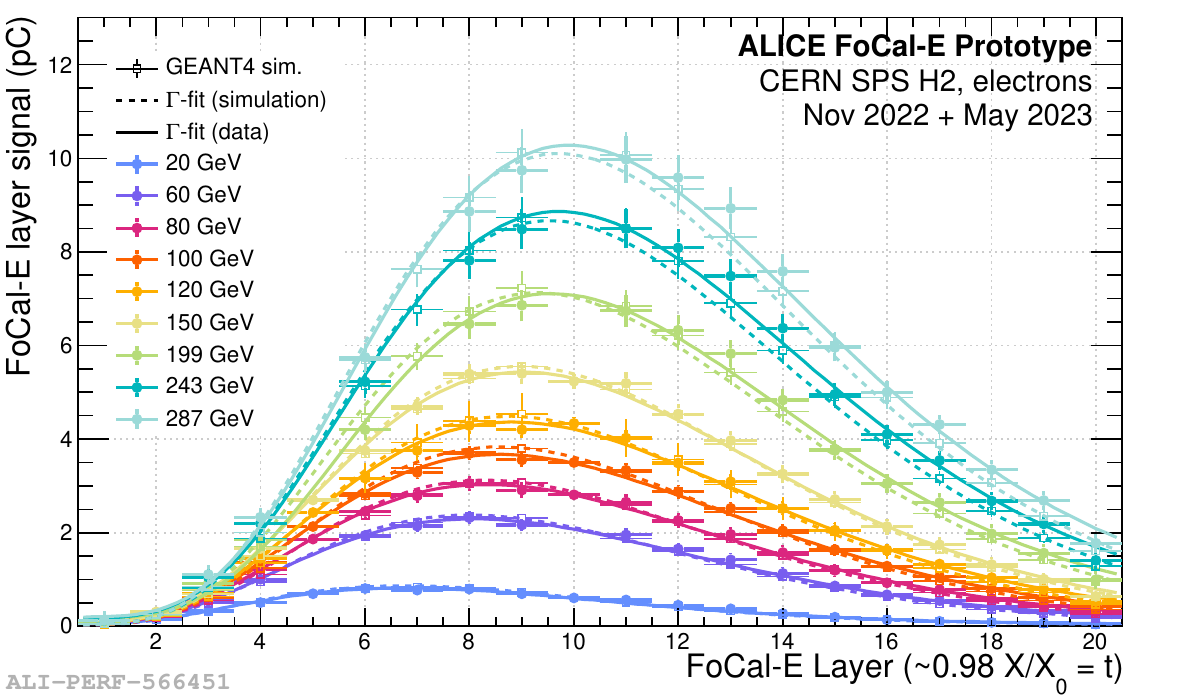}
    \end{subfigure}
    \caption{\textbf{Left:} Relative energy resolution for the FoCal-E pad layers. Results from May 2023 (with-Layer-7, blue markers) and combined May 2023 and November 2022 data (without-Layer-7, black markers) are compared to simulation (open markers), with respective fits. \textbf{Right:} Longitudinal shower profiles for 20–300 GeV electrons compared to \textsc{Geant4} simulations and fitted with a $\Gamma$-distribution. }
    \label{fig:tb_res}
\end{figure}

The performance of a full-length FoCal prototype has been tested and studied in an extensive test beam campaign at the CERN PS and SPS facilities from 2021 to 2023 \cite{tb}, with more test beams performed and scheduled also afterwards. Results from this campaign, using beams of electrons and hadrons up to 350 GeV in energy, have demonstrated the capabilities of FoCal in providing an energy resolution within the physics requirement of about 5\%, with a resolution of less than 3\% observed for high energies, as shown in Figure \ref{fig:tb_res} (left) for the FoCal-E prototype. The measured results are well described by simulations. For FoCal-H, resolution is below 15\% at high energies. In Figure \ref{fig:tb_res} (right), the longitudinal shower profile in FoCal-E showcases how the shower evolves as it passes through the twenty pad and pixel layers of the detector, offering a 3D view of the shower. 

\end{document}